\documentclass[twocolumn]{aastex631} 

\usepackage{graphics,epsf}
\usepackage[utf8]{inputenc}
\usepackage{amsmath}                
\usepackage{amsfonts}               
\usepackage{amssymb}                
\usepackage{epsfig}                 
\usepackage{graphicx}               
\usepackage{float}
\usepackage{color}
\usepackage{multirow}               
\usepackage{hyperref}

\hypersetup{
    colorlinks=true,
    linkcolor=red,   
    urlcolor=cyan}

\usepackage[colorinlistoftodos]{todonotes}

\newcommand{\s}{{~\rm s}}

\newcommand{\g}{{~\rm g}}
\newcommand{\G}{{~\rm G}}
\newcommand{\K}{{~\rm K}}

\newcommand{\keV}{{~\rm keV}}


\begin{document}

\title{The main jet axis of the W49B supernova remnant}
\date{January 2025}

\author[0000-0003-0375-8987]{Noam Soker}
\affiliation{Department of Physics, Technion - Israel Institute of Technology, Haifa, 3200003, Israel; soker@physics.technion.ac.il; s.dmitry@campus.technion.ac.il}

\author[0000-0002-9444-9460]{Dmitry Shishkin}
\affiliation{Department of Physics, Technion - Israel Institute of Technology, Haifa, 3200003, Israel; soker@physics.technion.ac.il; s.dmitry@campus.technion.ac.il}

\begin{abstract}
We identify an axis connecting two opposite `ears' in the supernova remnant W49B and morphological signatures of three arcs around this axis that we claim are sections of full circum-jet rings. Based on recent identifications of morphological signatures of jets in core-collapse supernovae (CCSNe), including ejecta-rich axes, we re-examine images of W49B and identify a heavy element-rich protrusion (ear) as a jet-inflated structure. We identify the opposite ear and a clump at its tip as the signature of the opposite jets. 
The line connecting the two clumps at the tips of the two opposite ears forms the main jet axis of W49B. We compare the three arcs around the main jet axis in W49B to the circum-jet rings of the jets in the Cygnus A galaxy and deduce that these arcs are sections of full circum-jet rings in W49B. In W49B, the jets are long gone, as in some planetary nebulae with circum-jet rings. Identifying the main jet axis is incompatible with a type Ia supernova. It leaves two possibilities: that jets exploded W49B as a CCSN, i.e., the jittering jets explosion mechanism where the pair of jets we identify is one of many that exploded the star, or that the explosion was a common envelope jet supernova with a thermonuclear outburst, i.e., both the pair of jets and thermonuclear outburst exploded the core of a red supergiant star as a pre-existing neutron star tidally destroyed it.  
\end{abstract}

\section{Introduction}
\label{sec:Introduction}

The supernova remnant (SNR) W49B  (G43.3–0.2; \citealt{Westerhout1958}) is a puzzling SNR concerning, among others, its explosion process and symmetry axis. The dynamically estimated age of W49B is 4-6 kyr (\citealt{Hwangetal2000,  ZhouVink2018}). Several studies presented its thermal and emission properties (e.g., \citealt{Ozawaetal2009, Patnaudeetal2015, HESScollaboration2018, Tanaka2018, Yamaguchietal2018, Liuetal2019, Sanoetal2021, Siegeletal2021, Suzukietal2024}), including images of W49B (e.g., \citealt{Lacyetal2001, Leeetal2019, Leeetal2020, Castellettietal2021}), and its properties. 
\citet{Lopezetal2009} studied in detail the X-ray morphology, and found the iron to have a different distribution from sulphur and silicon.
\citet{ZhouXetal2011} simulated the interaction of W49B with its inhomogeneous circumstellar and interstellar medium and claim that mixing of relatively cold medium gas with the hot shocked ejecta and the rapid adiabatic expansion of the ejecta explains the presence of overionized plasma, which was later strengthened by X-ray observations (Chandra: \citealt{Lopezetal2013b}; NuSTAR: \citealt{Yamaguchietal2018}). 
On the other hand, \citet{HollandAshfordetal2020} found the overionization to increase from east to west, like \citet{Lopezetal2013b} found, and argued that cooling by thermal conduction can explain this overionization.
\citet{SunChen2020} studied W49B and its overionized ejecta, and concluded that the ejecta's metal abundance ratios are compatible with a CCSN of a star with an initial mass of $<15 M_\odot$, except for a high Mn/Fe. \citet{Sanoetal2021} found that CO clouds at low velocity show a good spatial correspondence to the X-ray and synchrotron structure of the SNR, and \citet{Zhouetal2022} found strong interaction with a cloud. \citet{Zhuetal2014} studied the molecular clouds in the W49B environment and suggested that the warm dust comes from the evaporation of clouds interacting with W49B. The complicated structure of W49B and its environment promoted suggestions of several scenarios for its formation.

The three progenitors that the literature discusses are thermonuclear explosion as a type Ia supernova  (SN Ia; e.g., \citealt{Hwangetal2000, ZhouVink2018, Siegeletal2020, Satoetal2024}), core-collapse supernova (CCSN; e.g.,   \citealt{Lopezetal2011, Lopezetal2013a, Yamaguchietal2014, Patnaudeetal2015}), and a common envelope jets supernova with thermonuclear outburst \citep{GrichenerSoker2023}. Some studies find W49B to be a peculiar remnant that does not fit any scenario well (e.g., \citealt{Patnaudeetal2015, Siegeletal2020}). \citet{Sawadaetal2025} argue that the Fe-group ejecta mass ratios might result from either an SN Ia or a CCSN. \citet{Satoetal2024} claim that their determined titanium abundance excludes almost all hypernova/jet-driven supernova models. We note that they refer to an explosion driven by a fixed-axis jet that requires rapid pre-collapse core rotation (e.g., \citealt{Khokhlovetal1999, Leungetal2023}); they do not refer to nor exclude the jittering jet explosion mechanism (JJEM) of CCSNe (for recent simulations of the JJEM see \citealt{Braudoetal2025}). 

There is also a debate on the direction of the jet axis of W49B, with the two views orthogonal to each other. One group takes the jet axis to be the narrow high concentration of iron in the centre, in the general east-west direction (e.g., \citealt{Micelietal2008, Micelietal2010, Lopezetal2011, Lopezetal2013a, GonzalezCasanovaetal2014}), as the axis of the barrel-shaped morphology that \citep{Keohaneetal2007} suggest for W49B. The other view is that the general symmetry axis, the main jet axis, is more or less in the north-south direction, as \citet{BearSoker2017PNSNR} defined it and a few papers adopted (\citealt{Akashietal2018, Siegeletal2020, GrichenerSoker2023}). In this study, we strengthen the latter view. 

Recent new identifications of jet axes in several CCSN remnants (CCSNRs; e.g., \citealt{SokerShishkin2024Vela}; for a review, see \citealt{Soker2024}) and finding similarities with some jet-shaped morphologies in planetary nebulae (PNe; e.g., \cite{Soker2024Galax, Bearetal2025}) and cooling flow clusters (\citealt{Soker2024CF, Soker2024Galax}), motivate us to re-examine images of SNR W49B and to search for the main jets axis; we do this in Section \ref{sec:MainAxis}. In section \ref{sec:Comparison}, we compare the morphology of W49B with jet-shaped objects to strengthen our identification of the main jet axis. We summarize in Section \ref{sec:Summary} with a discussion on the possible origins of SNR W49B.    

\section{The main jet axis of W49B}
\label{sec:MainAxis}

\begin{figure*}
\begin{center}
\includegraphics[trim=0cm 0cm 0cm 0cm, clip, scale=0.70]{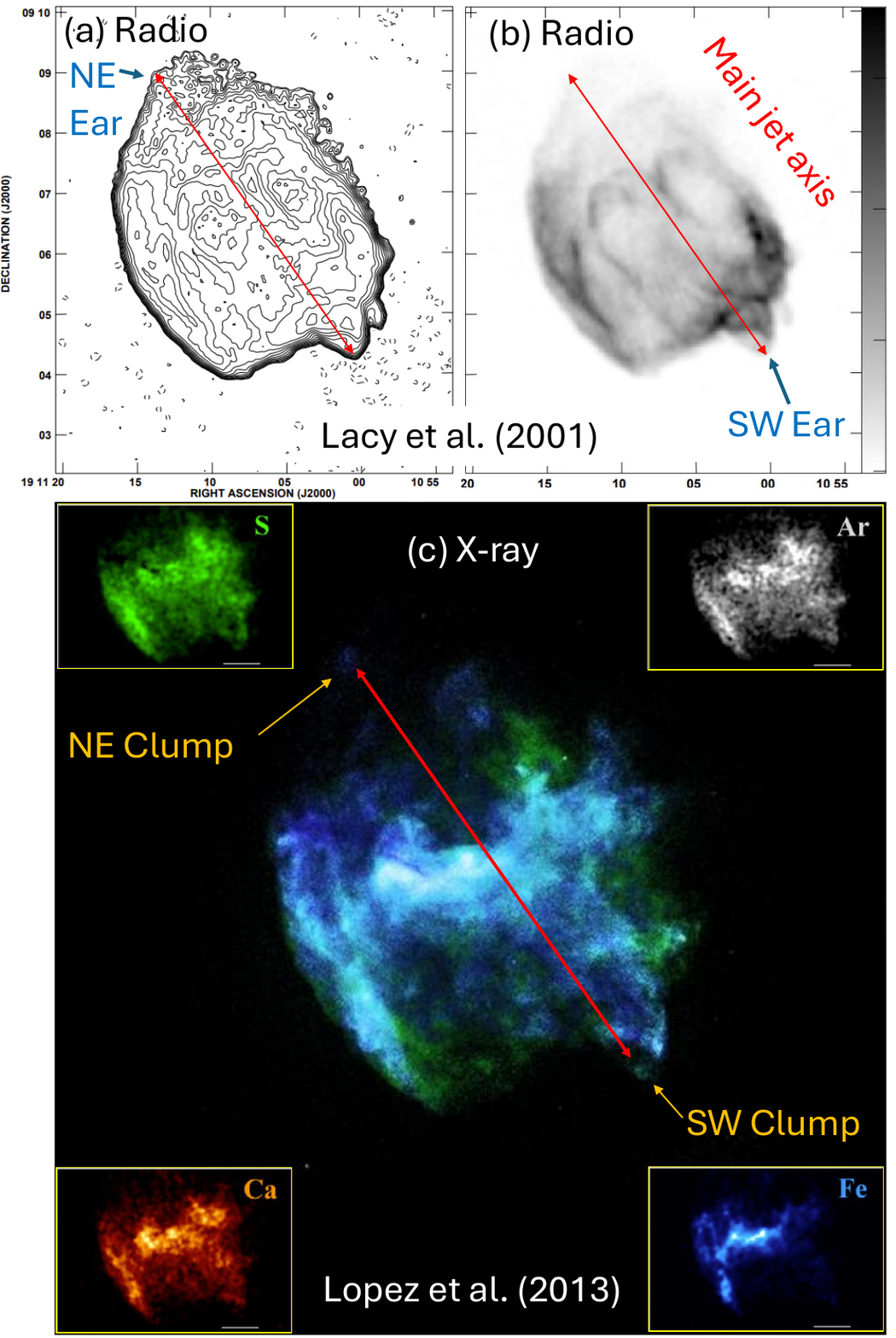} 
\caption{Panels a and b are radio images of SNR W49B from \citet{Lacyetal2001}, contours, and linear gray scale from 0 to 175 Jy/beam, respectively. Panel c is an X-ray image from Chandra ({\url{https://chandra.harvard.edu/photo/2013/w49b/}}; credit NASA/CXC/MIT/\citealt{Lopezetal2013a}). Inset images are from \citet{Lopezetal2013a} with the scale bar marking $1^\prime$. We labelled our proposed main jet axis with a double-headed red arrow between the two ears and the clump at the tip of each ear. }
\label{fig:W49Radio}
\end{center}
\end{figure*}

We start with radio images from \citet{Lacyetal2001} that we present in panels a and b in Figure \ref{fig:W49Radio} . The radio image shows a prominent ear in the southwest and a larger and less prominent one in the northeast, as we mark on panels a and b. Panel c is an X-ray image with two clumps at the tips of the ears, which we mark by NE and SW clumps. The four insets are images from \citet{Lopezetal2013a} that present the morphology of four elements as indicated. The double-headed red arrow has the exact location in all panels. 

\begin{figure*}
\begin{center}
\includegraphics[trim=0cm 0cm 0cm 0cm, clip, scale=0.70]{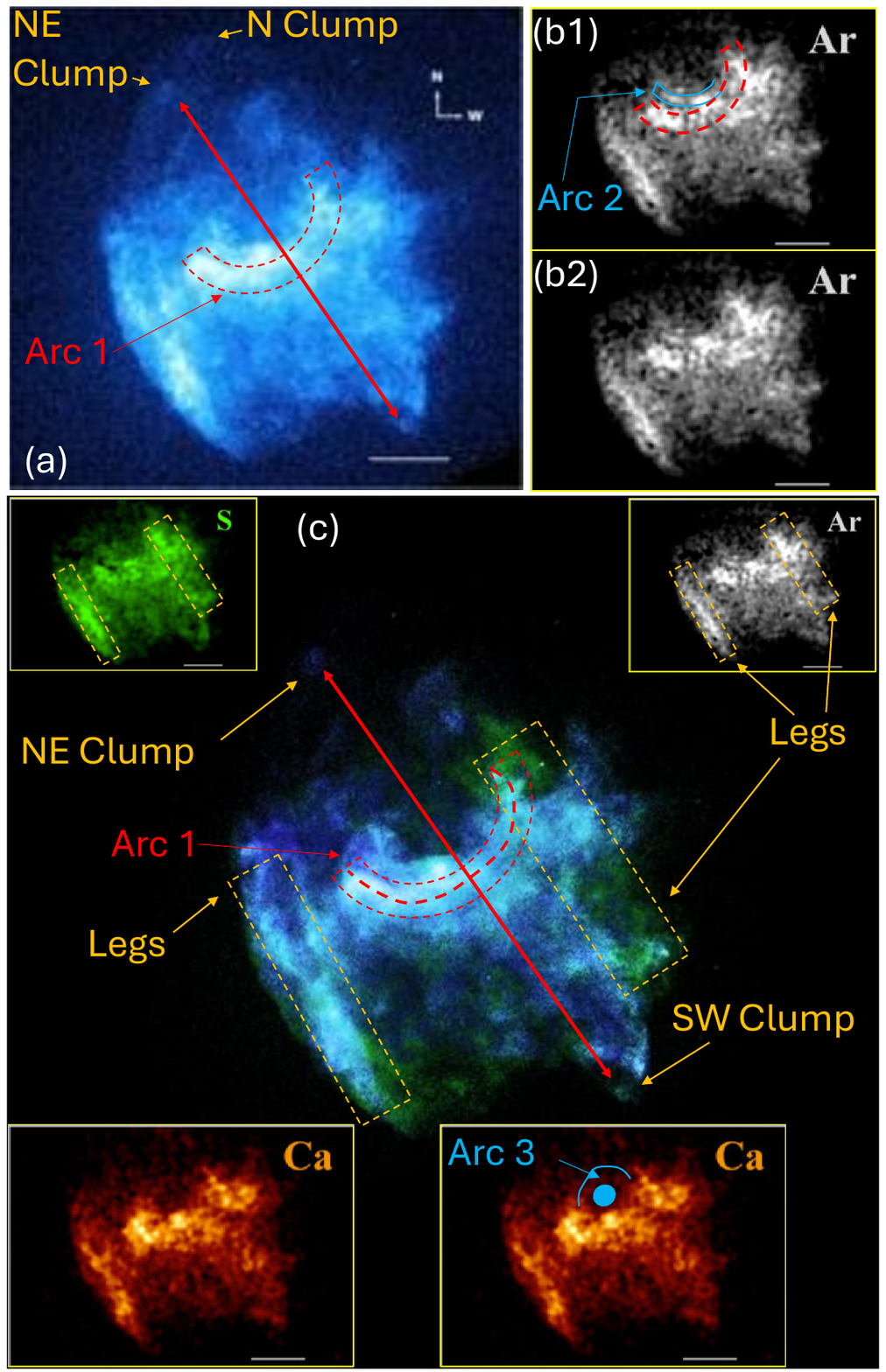} 
\caption{X-ray images of SNR W49B adapted from \citet{Lopezetal2013a} with our marks of morphological features we identify.  (a) A Chandra $0.5 - 8.0 \keV$ raw X-ray image. We marked our identification of the main jet axis, and an arc (dashed red line) we suggest is a fraction of a circum-jet ring. (b1+b2) Enlargement of the argon map (inset of panel c) to allow comparison of an image with and without our marks of two arcs. (c) Similar to panel c in Figure \ref{fig:W49Radio}, with the addition of marks as indicated. The two calcium panels allow comparison without our marks.}
\label{fig:W49Rings}
\end{center}
\end{figure*}

Figure \ref{fig:W49Rings} presents X-ray images adapted from \citet{Lopezetal2013a}. We identify arcs around the main jet axis; we will argue in Section \ref{sec:Comparison} that these are fractions of circum-jet rings the jets shaped when they were active at the explosion.  We identify three arcs as we mark on the different panels. The boundaries we draw partially go through faint zones around the bright arcs. The bright arcs are not complete, nor are the faint zones bounding the arcs. Nonetheless, we think the images do reveal these three arcs.  

\citet{BearSoker2017PNSNR} defined an `H-shaped' structure composed of the two legs, which are the bright regions on the northwest and southeast as we mark on Figure \ref{fig:W49Rings}, and the bright iron structure connecting them. The legs are the projection of the axisymmetrical sides of a barrel-shaped structure. 
Based on morphologies of H-shaped planetary nebulae, they assume the axis of the two opposite jets in W49B is parallel to the legs but did not use observed morphological features on the jet axis. \citet{BearSoker2017PNSNR} speculated on the location of the jet axis to be more or less parallel to the legs and approximately at an equal distance between the two legs.

We use the two bright clumps and the ears to locate the main jet axis somewhat to the northwest and at $4^\circ$ counterclockwise to the axis that \citet{BearSoker2017PNSNR} speculate at. It is more or less perpendicular to the jet axis that some other studies assume (e.g., \citealt{Keohaneetal2007, Micelietal2008}; see Section \ref{sec:Introduction}).

A word of caution is in place here. We assume that the ears are shaped by jets that are part of the explosion process. However, in principle,  ears can be formed in the circumstellar material into which the explosion occurs, not by jets in the explosion. This might be the case with type Ia SN remnants that have ears. The explosions of type Ia SNe are not expected to include jets, hence the ears are already shaped in the medium around the explosion (e.g., \citealt{Soker2024RAA} for type Ia SNR G1.9+0.3). If the ears are shaped in the circumstellar material, no signatures of the symmetry axis the ears define exist near the centre of he explosion. In W49B, there is a non-spherical mass distribution near the centre that, as we claim below, has the same symmetry as the line connecting the two ears. We therefore argue that the ears in W49B are shaped by jets.

Like \citet{BearSoker2017PNSNR}, we argue that two opposite jets shaped the barrel-like structure, as observed in some planetary nebulae and as numerical simulations show (e.g., \citealt{Akashietal2018}). The hot gas in the galaxy M84 possesses a prominent H-shaped structure shaped by the observed still-active active galactic nucleus (AGN) jets (e.g., \cite{Bambicetal2023}\footnote{ \url{https://chandra.harvard.edu/blog/node/843}}); in M84 the jets are parallel to the leg, but the jets' axis is displaced and much closer to one leg than the other. What \citet{BearSoker2017PNSNR} attributed to a compressed gas in the equatorial plane (the plane between the two jet-inflated lobes), we attribute to Arc 1 that we mark on Figure \ref{fig:W49Rings}; some gas might be due to gas compression by the jets in the equatorial plane. 

\citet{BearSoker2017PNSNR} assumed that the two jets were energetic to the degree that they opened the SNR W49B on the northeast and southwest sides (top and bottom of the barrel-shaped structure), and that is the reason for the absence of apparent morphological features along the jet axis, especially in the southwest. The small ear on the southwest and the small SW clump suggest that this jet had a small fraction of the total explosion energy. The conclusion is that the jets we identified as having shaped the ears did not supply the entire explosion energy. There are two possibilities with which our results are compatible. (1) More pairs of jets powered the explosion in case W49B is a descendant of a CCSN, i.e., the explosion was due to the JJEM. (2) Thermonuclear energy supplied the rest of the explosion energy, as in the common envelope jets supernova with thermonuclear outburst model suggested by \citet{GrichenerSoker2023}.

\section{Comparison to other objects with circum-jet rings}
\label{sec:Comparison}

The main new claim of this study is that some of the dense (bright) ejecta material near the centre of W49B is not the bar of an H-shaped structure nor a jet but instead arcs that are part of circum-jet rings. We compare W49B with other astrophysical objects to strengthen this new claim. 

Aiming to explain W49B structure, \citet{GonzalezCasanovaetal2014} simulated the interaction of one pair of energetic jets with ambient gas in the frame of the fixed-axis jet-driven explosion model of CCSNe (magnetorotational jet explosion model; see \citealt{RamirezRuizMacFadyen2010} for another simulation of fixed-axis jets with comparison to W49B). Most relevant to us is that \citet{GonzalezCasanovaetal2014} obtained the appearance of a few rings around the jet axis; however, they did not identify the circum-jet rings we define here in W49B. We will use rings observed in other astrophysical objects instead of numerical simulations. 

\subsection{SNR 0540-69.3}
\label{subsec:SNR0540693}

\begin{figure}[ht!]
\includegraphics[trim=0.0cm 16.1cm 0.0cm 2.5cm ,clip, scale=0.79]{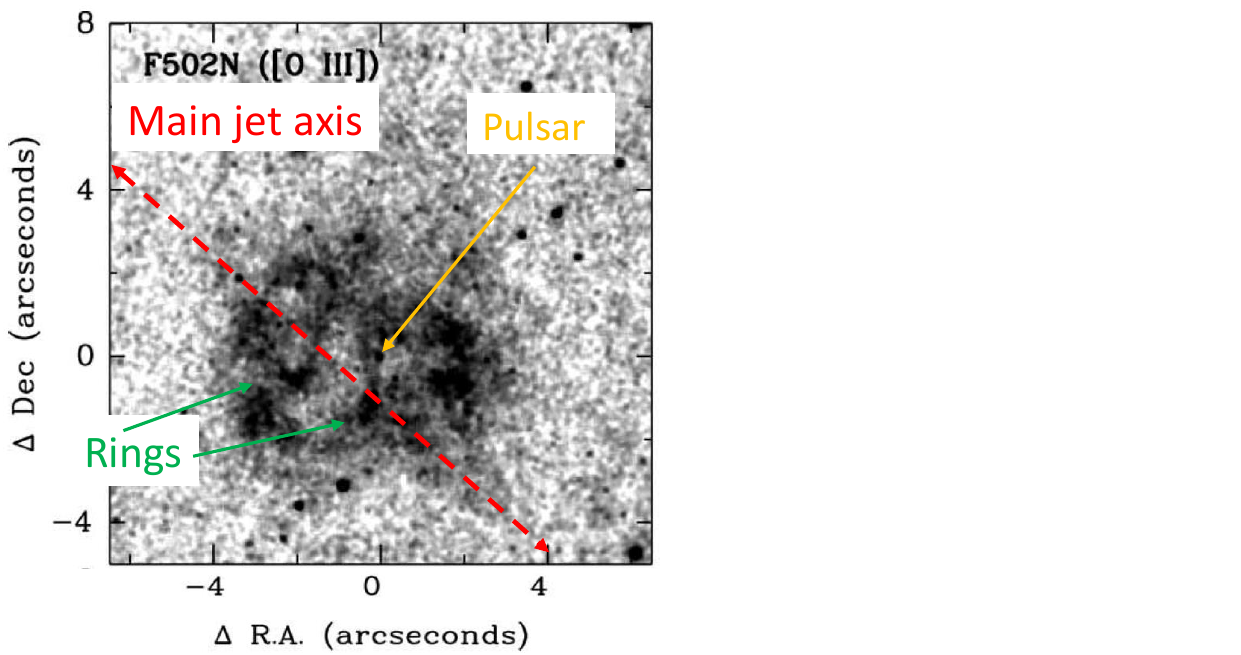}
\caption{An HST image of SNR 0540-69.3 adapted from \citet{Morseetal2006}. The red double-headed arrow marks the main jet axis according to \citet{Soker2022SNR0540}. We also marked two circum-jet rings that we identify in this study.}
\label{fig:MorseFig}
\end{figure}

Figure \ref{fig:MorseFig} presents an HST image of SNR 0540-69.3 adapted from  \citet{Morseetal2006}. 
Based on the slit spectroscopy by \citet{Larssonetal2021}, \citet{Soker2022SNR0540} identified a point-symmetric morphology in a plane along the line of sight (not shown here), and the main jet axis that we mark with the red-double-headed arrow in Figure \ref{fig:MorseFig}. We mark two circum-jet rings that we identify in the HST image. 
We see some similarities between the circum-jet rings around the main-jet axis in SNR 0540-69.3 and the arcs we argue are part of circum-jet rings in SNR W49B.  

\subsection{Planetary nebulae}
\label{subsec:PNe}

We present two PNe with prominent circum-jet rings. Since the jets in PNe are no longer active, their signatures are clumps along the polar axis isolated from the centre and the left-over circum-jet rings. 

\begin{figure}[ht!]
\includegraphics[trim=0.0cm 13.cm 0.0cm 0.0cm ,clip, scale=0.52]{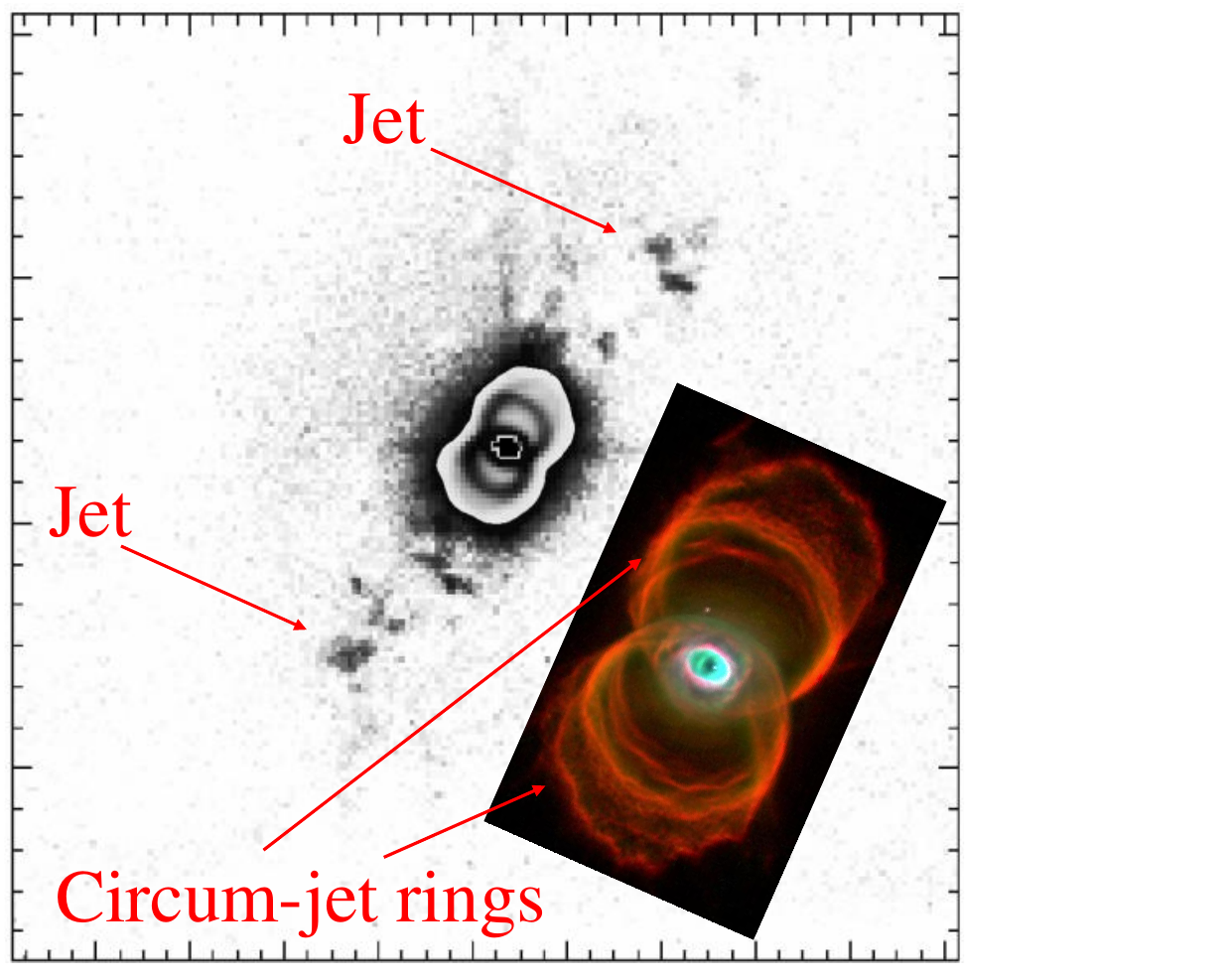}
\caption{An image of the PN MyCn 18 adapted from \citet{OConnoretal2000}. The inset is an HST image from the HST site adapted from \citet{Sahaietal1999MyCn18} and resolves the circum-jet rings in the hourglass structure.}
\label{fig:MyCn18}
\end{figure}

Figure \ref{fig:MyCn18} presents the bipolar PN MyCN 18. The inset (based on \citealt{Sahaietal1999MyCn18}) is an HST image of the inner hourglass structure. It possesses several rings. The large image includes the regions along the polar axis that display clumps formed by no-long-active jets (e.g., \citealt{OConnoretal2000}). Therefore, the rings that compose the hourglass structure are circum-jet rings.  

\begin{figure}[ht!]
\includegraphics[trim=0.0cm 10.cm 0.0cm 0.0cm ,clip, scale=0.60]{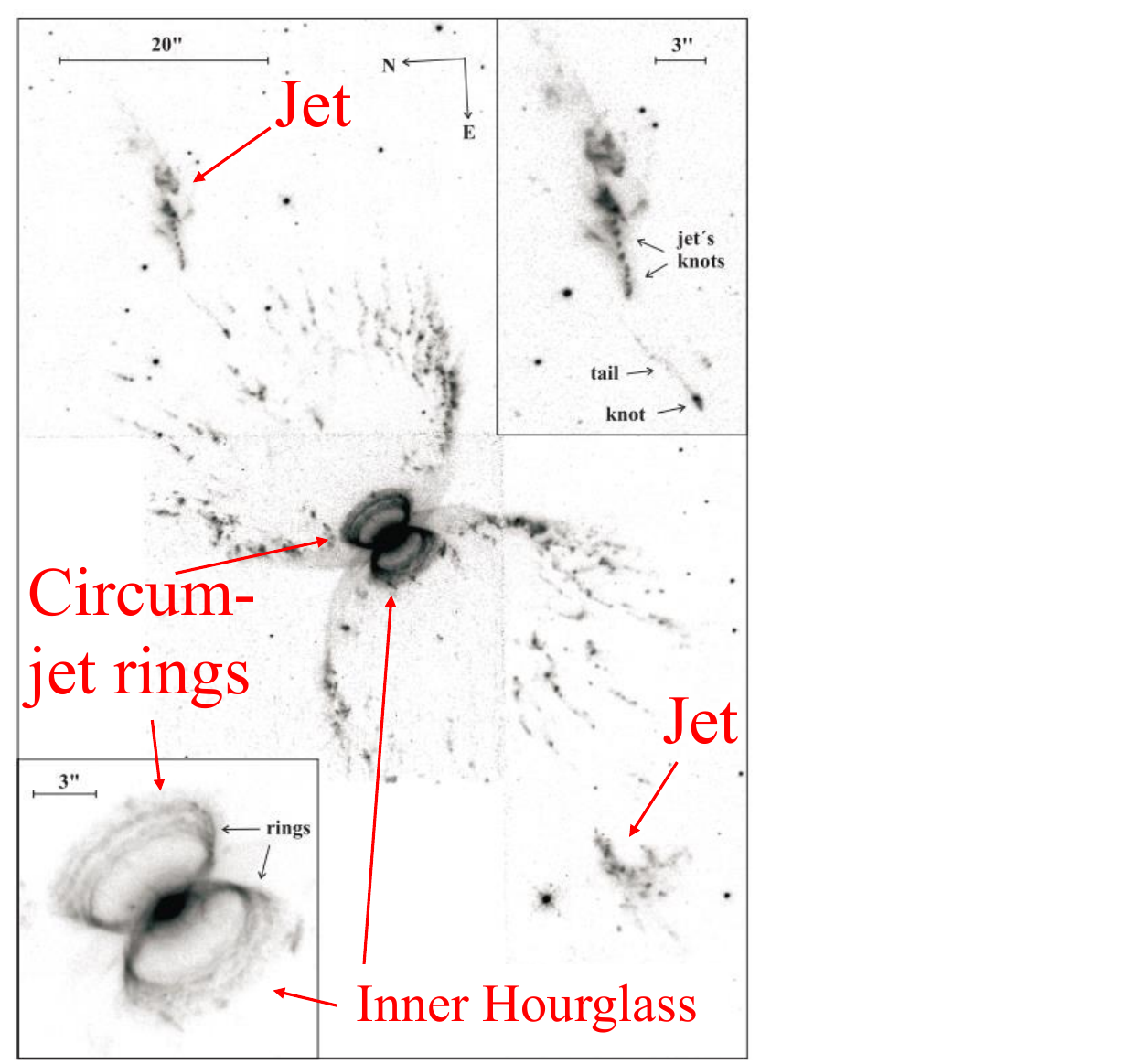}
\caption{[N II] HST image of the PN Hen 2-104 adapted from \citet{Corradietal2001}; the black marks of jets and rings are from their paper. We added red marks to emphasize the circum-jet rings in the inner hourglass.}
\label{fig:Hen2104}
\end{figure}

Another PN with signatures of jets and circum-jet rings is Hen 2-104 (the `Southern Crab Nebula'). 
Figure \ref{fig:Hen2104} presents an image adapted from \citet{Corradietal2001}. The black marks of jets and rings are in the original study of \citet{Corradietal2001}. We added marks in red to emphasize the jets and the circum-jet rings. 
The three-dimensional reconstruction of Hen 2-104 structure by \citet{Clyne2015} shows that three rings compose each lobe of the inner hourglass structure. 

These two PNe demonstrate the presence of circum-jet rings in objects where the jets are not active anymore. In PNe, such rings cannot be attributed to explosion processes or the energy of radioactive decay (called nickel bubbles in CCSNe). Their mirror and axial symmetries imply that they do not result from instabilities. 

\subsection{Clusters of galaxies}
\label{subsec:Clusters}

Cygnus A is a galaxy with active AGN jets and is another example of jet launching with circum-jet rings.
\citet{Soker2024CF} compared the rings of SNR 0540-69.3 (Figure \ref{fig:MorseFig}) to those of Cygnus A to strengthen the identification of the main jet axis of SNR 0540-69.3. We argue for a similarity between the bright segments of the rings in Cygnus A and the arcs in the SNR W49B. 

\begin{figure}
\begin{center}
\includegraphics[trim=1.5cm 4.6cm 1.5cm 6cm ,clip, width=0.52\textwidth]{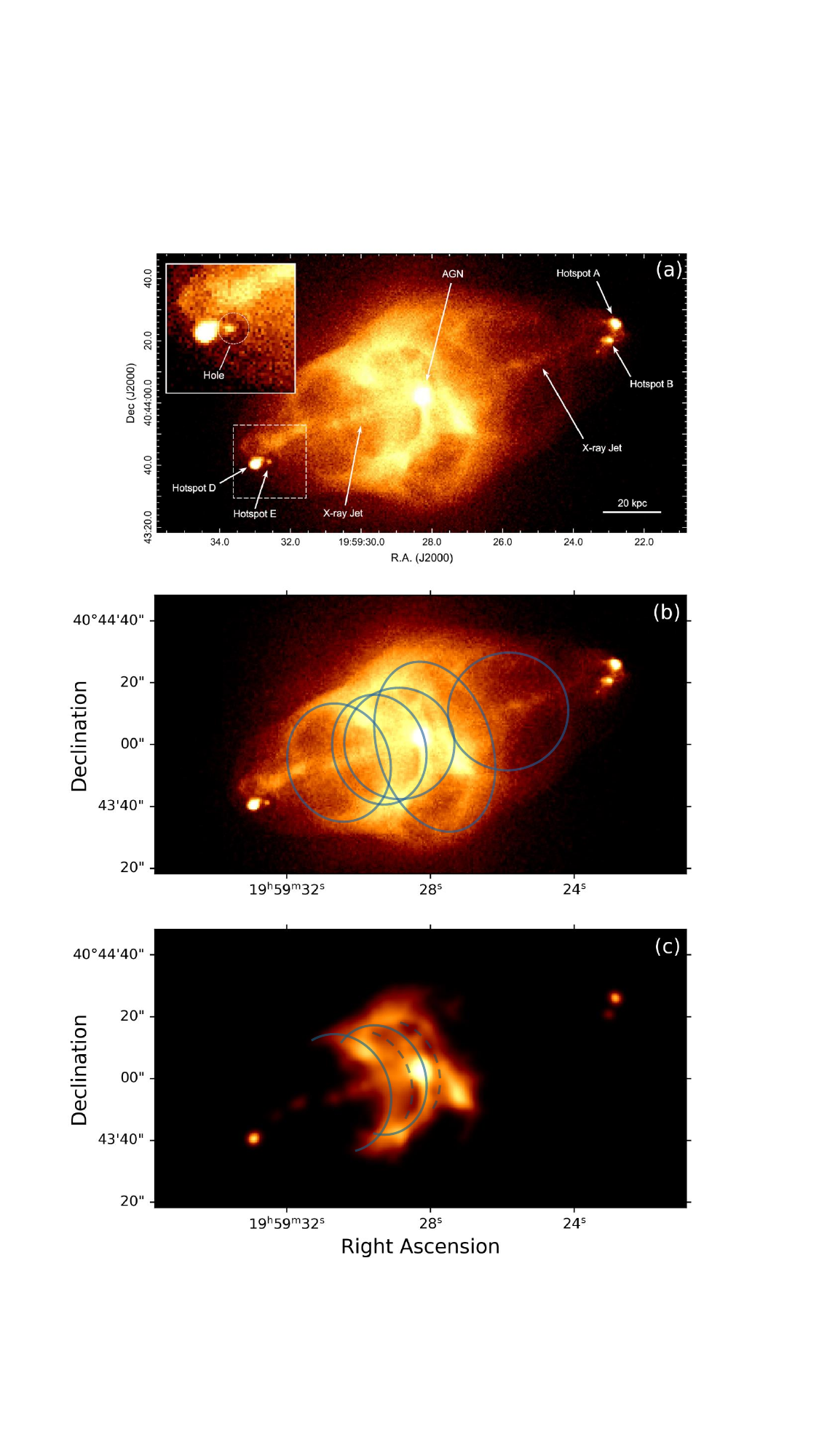} 
\end{center}
\caption{An X-ray image of the cluster of galaxies Cygnus A (0.5-7 keV; adapted from \citealt{Sniosetal2020}). (a) Figure 1 from \citet{Sniosetal2020} with their original marks. The horizontal bar is a scale of 20~kpc. (b) The same X-ray image, with pale ellipses on top of several possible rings (see also \citealt{Soker2024CF} for comparison with SNR 0540-69.3). (c) Our processed image of Cygnus A with reduced sensitivity and resolution (reduced version). Two rings denoted in panel b appear as arcs in panel c. Dashed arcs denote faint regions between rings. Two opposite clumps (or hotspots) are the only visible parts of the jets in this degraded image. }
\label{Fig:CygnusA} 
\end{figure}

Figure \ref{Fig:CygnusA} presents an X-ray image of the galaxy Cygnus A adapted from \citet{Sniosetal2020}. We identify several rings, which we denote with blue circles in Figure \ref{Fig:CygnusA} panel b. The centres of most of these circles align with the X-ray jet. We note that the rings are not uniform in intensity; in each ring, some segments are brighter than others, and the bright segments form arcs. 

To strengthen the comparison of the circum-jet rings of SNR W49B to those in Cygnus A, we degrade the $2Ms$ Chandra X-ray image of Cygnus A from \citet{Sniosetal2020}. We use Figures 1 and 2 from the \citet{Sniosetal2020}, which we then convert to a scalar image. 
To match the existing W49B observations, we perform two adaptations to this scalar image.  (1) We truncate the lower $50\%$ brightness pixels to correct for the 10 times shorter observation time of W49B \citep{Lopezetal2013b}. (2) We apply Gaussian smoothing with a 2-pixel scale to correct for the 4 times larger field of view of W49B compared to Cygnus A. 
We present the degraded image of Cygnus A in the lower panel of Figure \ref{Fig:CygnusA}. Only a fraction of the brightest rings are visible in this processed image, forming arcs. We mark the less bright regions between the bright arc segments with dashed lines.

Another interesting feature of the degraded Cygnus A's image in panel c of Figure \ref{Fig:CygnusA} is that the only parts visible from the jets are two opposite clumps. Such clumps can testify to the activity of jets, at present, as in Cygnus A, or in the past, as in the planetary nebulae we present in Figures \ref{fig:MyCn18} and \ref{fig:Hen2104}.   

\begin{figure}
\begin{center}
\includegraphics[trim=0.42cm 5.5cm 0.5cm 6cm, clip, width=0.51\textwidth]{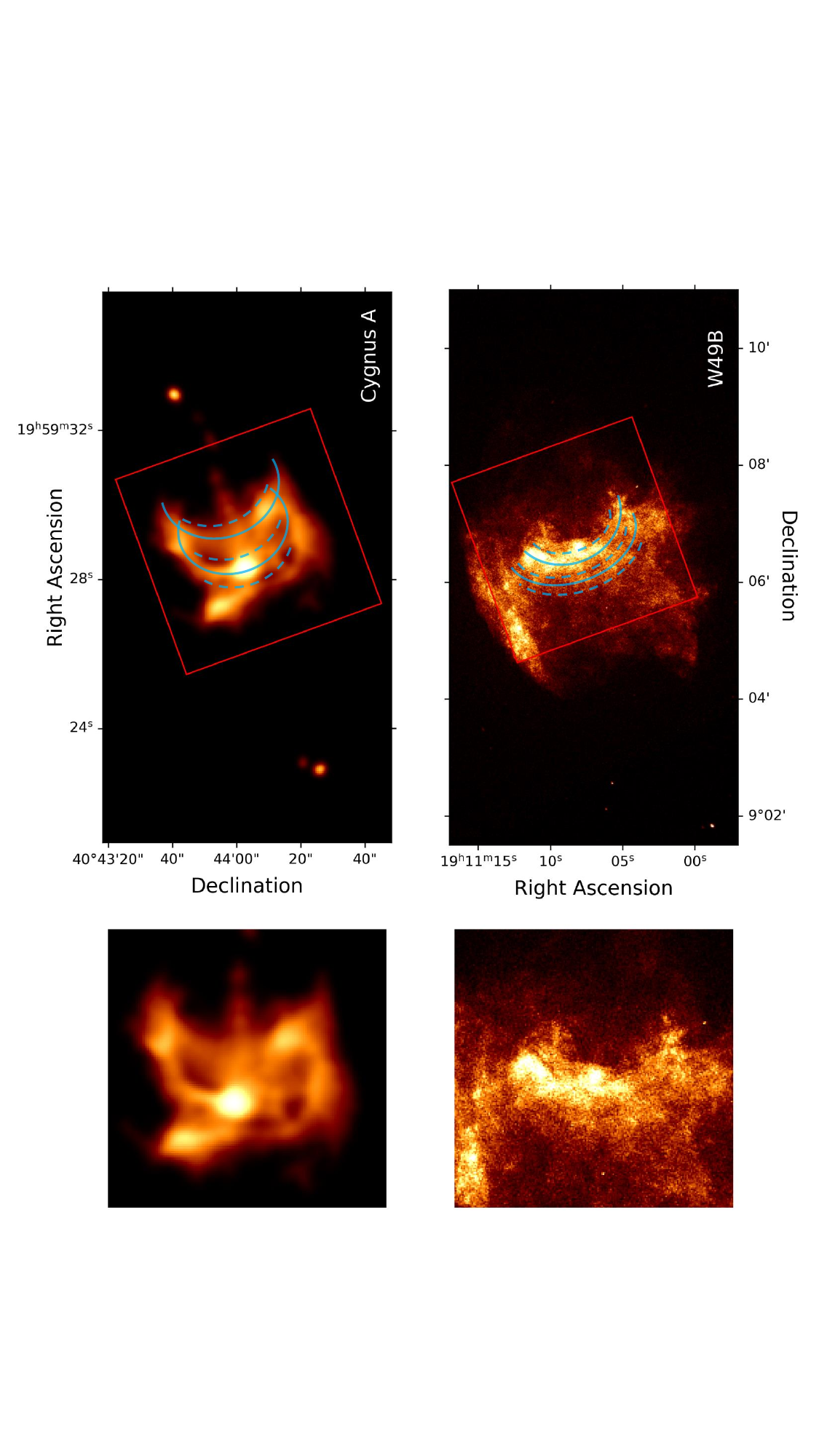} 
\end{center}
\caption{A comparison between an X-ray image of W49B and panel (c) of Figure \ref{Fig:CygnusA} rotated $90^\circ$ clockwise. Upper panels: Side-by-side comparison of X-ray images of Cygnus A and W49B. Solid arcs denote parts of rings, and dashed arcs denote regions of lower brightness between (or adjacent to) rings. Red rectangles denote the area of interest displayed in the lower panels. Lower panels: Zoom-in on the areas of interest, showing the ringed inner structure of Cygnus A within the vicinity of the jet-launching AGN (left) and X-ray bright region of W49B where we identify a similar ringed structure (right).}
\label{fig:CygnusA_W49B} 
\end{figure}

To illustrate the similarity with W49B, we rotate the image in panel c of Figure \ref{Fig:CygnusA} in the plane of the sky by $90$ degrees clockwise, such that right ascension and declination are now the vertical and horizontal axis respectively.
We present our reduced version and a comparison image of W49B in Figure~\ref{fig:CygnusA_W49B}.

We identify a similar structure in the reduced image of Cygnus A and the X-ray image of W49B, consisting of arcs which we attribute to being partially visible rings (solid lines in Figures \ref{Fig:CygnusA} and \ref{fig:CygnusA_W49B}), and lower brightness gaps between arcs (dashed lines in Figure \ref{Fig:CygnusA}).
This comparison further strengthens the similarity between an established "jetted" system (Cygnus A) structure and the observed inner structure SNR W49B.

\section{The formation of circum jet rings}
\label{sec:Rings}

In principle, three classes of interactions might form circum-jet rings: We discuss them here, but leave the simulations to future studies.

(1) \textit{Continuous jet propagation into a shell.}
\citet{AkashiSoker2016Ring} simulated the interaction of a pair of opposite jets with a shell, followed by a fast low-density wind. They showed that this type of interaction leaves a circumjet ring where each jet interacts with the shell. 
Their aim was planetary nebulae. The core with which the jets interact in CCSNe does not have a dense shell. Therefore, this scenario does not apply to CCSNe (unless the rings are from jets interacting with a circumstellar shell; this is irrelevant to the rings in W49B because the rings are near the centre).  

(2) \textit{Several jet-launching episodes in the same direction into a smooth environment.} Let us then consider the case of two (or more) jet-launching episodes in the same direction into a smooth core. 
The first jet inflates a bubble that compresses a cap at its front. The next jet along the same direction interacts with this denser than the environment cap, and the interaction forms a ring, as in the simulation by \citet{AkashiSoker2016Ring}.  

(3) \textit{Instabilities.} The interaction between a jet and a surrounding gas is prone to Kelvin–Helmholtz instabilities. The shocked hot low-density gas forms a chain of vortices, and between these vortices there is dense gas around the jet (e.g., \citealt{RefaelovichSoker2012}).
\citet{GonzalezCasanovaetal2014} conducted simulations of jets to explain the structure of W49B. They obtained rings from instabilities in the interface between the jet inflated bubble and the surrounding medium. They did not relate these rings to what we identify as rings. More simulations are needed to check whether instabilities can explain the observed rings. 

Considering the thick torus of the rings in W49B, we inclined towards the second scenario above. However, only dedicated three-dimensional simulations can determine the outcome of such interactions.

\section{Summary}
\label{sec:Summary}

We addressed the question of the location of the main jet axis in the enigmatic SNR W49B. To that end, we take the two radio ears (Figure \ref{fig:W49Radio}) to be two structures inflated by two opposite jets, and the line connecting the ears is the main jet axis. Two X-ray clumps on the tips of the ears are compatible with this identification because dense clumps are remnants of ears in many other jet-shaped objects, as in planetary nebulae (Figures \ref{fig:MyCn18} and \ref{fig:Hen2104}). 
We identified the bright zones at the centre of SNR W49B as arcs, which we argue are part of complete rings (Figure \ref{fig:W49Rings}). Circum-jet rings exist in many jet-shaped objects, like SNR 0540-69.3 (Figure \ref{fig:MorseFig}), planetary nebulae (Figures \ref{fig:MyCn18} and \ref{fig:Hen2104}), and AGN jets. Cygnus A's cooling flow cluster jets shaped several circum-jet rings (Figure \ref{Fig:CygnusA}). We degraded the X-ray image of Cygnus A by removing fainter zones and reducing its resolution. The outcome is an image where only segments of the rings are visible and the bright ends of the two jets (clumps). Namely, an image of arcs around the axis connecting the two opposite clumps (panel c of Figure \ref{Fig:CygnusA}). We argued that the X-ray image of SNR W49B shares many properties with the degraded image of Cygnus A (Figure \ref{fig:CygnusA_W49B}). Since Cygnus A jets are visible and still active, this similarity supports our claim for circum-jet rings in SNR W49B, where jets are long gone, and for the location of the W49B main jet axis. 

The direction of the main jet axis we claim for, is $4^\circ$ counterclockwise to the axis that \citet{BearSoker2017PNSNR} argued for. The more significant difference is in the jets' energy.
\citet{BearSoker2017PNSNR} speculated that one pair of jets was energetic enough to shape the entire structure of W49B, including the legs (Figure \ref{fig:W49Rings}). In their model, the jets broke out from the main part of the ejecta to the northeast and southwest. 
In this study, we argue for a pair of jets that shape the ears but do not break out from the ejecta. This pair is not energetic enough to shape the entire ejecta. The additional explosion energy might come from more jets or nuclear burning. 
If SNR W49B is a CCSN remnant, many more pairs of jets can supply the extra energy in the frame of the JJEM. In that case, one or two pairs might be along the direction that \citet{BearSoker2017PNSNR} speculated. 
If SNR W49B is a remnant of a common envelope jets supernova, then nuclear burning of the destroyed core can supply the extra energy (for the different scenarios, see Section \ref{sec:Introduction}). 

Our identification of the signatures of jet activity, i.e., the shaping of ears and circum-jet rings, does not support claims that W49B is a remnant of SN Ia explosion.
We leave the comparison of the two plausible scenarios, CCSN explosion in the frame of the JJEM or common envelope jets supernova, for the future. 

\section*{Acknowledgements}
We thank an anonymous referee for helpful comments.

\paragraph{Funding Statement}
A grant from the Asher Space Research Institute in the Technion supported this study.

\end{document}